\documentstyle[twocolumn,eqsecnum,aps,epsfig]{revtex}

\begin{document}

\title{Evidence of early 
  multi-strange  hadron freeze-out in high energy nuclear collisions }

\author{H. van Hecke$^{a}$, H. Sorge$^{b}$, and N. Xu$^{c}$}

\address{\em (a) MS H846, LANL, Los Alamos, NM 87545, USA}

\address{\em (b) Department of Physics, SUNY at Stony Brook, NY 11794, USA}

\address{\em (c) Nuclear Science Division, LBNL, Berkeley, CA 94720, USA}

\date{\today}
\maketitle
\begin{abstract}
Recently reported transverse momentum distributions of strange hadrons
produced in Pb(158AGeV) on Pb collisions and corresponding results
from the relativistic quantum molecular dynamics (RQMD) approach are
examined.  We argue that the experimental observations favor a
scenario in which multi-strange hadrons are formed and decouple from
the system rather early at large energy densities (around 1
GeV/fm$^3$).  The systematics of the strange and non-strange particle
spectra indicate that the observed transverse flow develops mainly in
the late hadronic stages of these reactions.
\end{abstract}
%
%
\narrowtext

 The purpose of the current and forthcoming heavy ion programs at the
  high-energy laboratories CERN (Switzerland) and Brookhaven National
  Laboratory (USA) is to probe strongly interacting matter under
  extreme conditions, i.e.  at high densities and temperatures. The
  central subject of these studies is the transition from the
  quark-gluon plasma to hadronic matter.  In the early phases of
  ultrarelativistic heavy ion collisions, when a hot, dense region is
  formed in the center of the reaction, there is copious production of
  up, down, and strange quarks.  Transverse expansion is driven by the
  numerous scatterings among the incoming and produced particles.  As
  the medium expands and cools, the quarks combine to form the hadrons
  that are eventually observed.

 In this Letter we are going to address the question whether data from
 heavy ion experiments allow one to make statements about the
 existence of strange hadrons in a medium of high energy density
 $\epsilon >$ 1GeV/fm$^3$. So far, experimental information about
 survival or ``melting'' of quark bound states was restricted to
 charmonium states only \cite{na50g}. On the other side, the strange
 quark mass is intermediate between charm and the very light flavors
 (up and down). Strangeness should in fact not be considered as an
 ``external probe'' of the hot medium like charm, as it may influence
 fundamental characteristics of the QCD phase transition itself,
 e.g. its order \cite{pw84}.  Furthermore, we argue in this Letter
 that the strange and non-strange hadron spectra encode information on
 the timing of flows and pressures.  It has been recognized for a long
 time \cite{vH83} that hadron momentum spectra are a valuable source
 of information on the collective flow developing in ultrarelativistic
 heavy ion collisions.  The flows may be related to bulk and transport
 properties in the ultra-dense matter like transient pressure and
 viscosities.  Presence of hydrodynamic behavior is expected at least
 for truly large colliding system such as Pb+Pb.  Concerning the
 search for the quark-gluon plasma one would like to identify the
 regions in energy density at which the Equation of State (EOS)
 softens, presumably due to the phase transition or a cross-over
 between hadronic and quark matter. Another topic of interest is to
 see the EOS becoming hard again at yet higher densities.
 Asymptotically, the EOS approaches the Stefan-Boltzmann limit
 $p=\epsilon /3$, where $p$ is the pressure, according to recent
 lattice calculations \cite{Ka95}.  Note, however, that the regime of
 perturbative quarks and gluons seems to be reached only at grand
 unification scales \cite{Kaj97}.  In the context of ultrarelativistic
 collisions the EOS dependence on energy and baryon density translates
 into a pressure dependence on time, because expansion dilutes the
 matter continuously.  Here hadrons with varying strangeness content
 play an important role. With vastly differing reaction rates in the
 medium, they decouple at different times from the evolving system.
 Pictorially speaking, we may employ these spectra to get a sequence
 of snapshots of the transverse flow present at each of the
 species-dependent decoupling times.

 On the experimental side, the predicted presence of strong radial
 transverse flow in the Pb(158AGeV) on Pb collisions (0.4 to 0.6 c)
 \cite{Sor95} has been deduced from the systematics of non-strange
 particle spectra already some time ago
 \cite{our_flow_paper,na44_coll_exp}.  The long awaited spectra of
 multiple strange hadrons $\Phi, \Xi,$ and $\Omega$, measured at
 mid-rapidity, were reported during the Quark Matter '97 conference
 \cite{tsukuba,na49qm97}.  It came somewhat as a surprise that the
 reported slopes of these multi-strange hadrons ($\Xi$'s and
 $\Omega$'s and possibly $\Phi $'s) are much softer than expected from
 the trends in the previously observed mass dependence of the slope
 parameters \cite{na44_coll_exp}.
 
 In this Letter we will argue that these deviations find their natural
 explanation in a reaction scenario in which the multi-strange hadrons
 freeze out early, before sizable transverse flow has been
 developed. The corollary statement is that the transverse flow is a
 phenomenon emerging only later in the Pb on Pb reactions.  In order
 to go beyond a qualitative interpretation we utilize a transport
 theoretical approach -- relativistic quantum molecular dynamics
 (RQMD) -- whose predictions agree well with the observations.
   
 Several scenarios for the unknown earlier stages in ultrarelativistic
 nucleus-nucleus are imaginable and have been put forward.  We will
 try to use the most recent information on the spectra to eliminate
 some of them.  Let us list some of the possible choices: 

\begin{itemize}
\item Is the expanding matter made of weakly interacting quarks and
 gluons, or of bound states (hadrons)?

\item Does the flow appear already at an early stage of the collision
\cite{Fe96} or rather later \cite{Sor97}?

\item Does freeze-out occur instantaneously \cite{BMSWX96} or
sequentially \cite{BRA95,Sor96}?
\end{itemize} 

 We do not believe that the existence of strong transverse flow in the
 Pb(158AGeV) on Pb reactions belongs to the unresolved questions. It
 is well-known that the difference between flow -- a space-momentum
 correlation -- and excitations purely in momentum space (temperature
 or random kicks \cite{satz}) shows up in observables which are
 sensitive to the phase-space densities of the finally emitted hadrons
 (like nucleon cluster formation \cite{Sor95} and HBT \cite{THP98}).
  
\vspace{-1.cm}
\begin{center}
\begin{figure}[h]
\centerline{\epsfig{figure=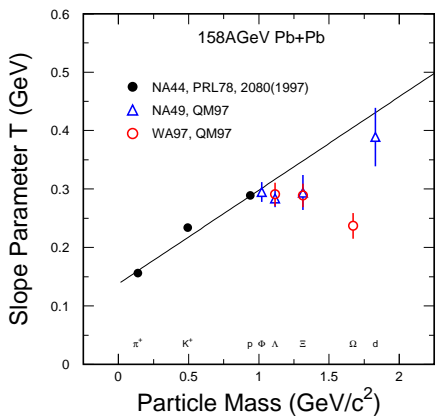,width=7.0cm}}
\caption{ Measured slope parameters as a function of
particle mass. Preliminary results of the slope parameters of strange
particles are shown with open symbols. 
The line represents the parameterization mentioned in the text.
 }
\end{figure}
\end{center}
\vspace{-0.5cm}

 Let us turn now to the recent data and interpret them in the light of
 these ideas.  A compilation of the experimental data is shown in
 Fig.~1.  The non-strange data may be parameterized by $T$=$ T_{fo} + m
 \langle \beta_t \rangle^2$, where $T$ the slope parameter at
 mid-rapidity \cite{fn_T} and $m$ the particle mass. The parameter
 values $ T_{fo} \approx $145 MeV, $\langle \beta_{t} \rangle = 0.4c$.
 Recent reports show that this trend is not only obeyed by $\pi^{+/-},
 p, \bar{p}$ and deuterons but also by hadrons containing one single
 strange quark \cite{na44_coll_exp,na49qm97}.  However, note the
 strong deviation of the multiple strange baryon ($\Xi^-$, $\Omega$)
 slopes from the general trends.  How much the $\Phi $ deviates from
 the general systematics is not yet settled \cite{Bor97}.

 We may draw three conclusions from the compilation of slope
 parameters.  The frequently employed picture -- one-fluid flow until
 break-up of the matter at a common freeze-out state -- is untenable
 in view of these data.  Such a picture results in a unique dependence
 of slope parameters on the particle mass, in clear contrast to the
 observations.  Note that feed-down corrections which could spoil the
 universal dependence are of relevance mostly for pions but not the
 ``heavy'' hadron sector (with masses close to and above 1 GeV).
 Furthermore, the striking difference between multi-strange hadron and
 the other hadron slopes points to dynamics in the hadronic stages
 that causes the non-universal pattern.  It is well-known from the
 early ideas of ``strangeness as a QGP signal'' \cite{KOC86} that
 strange flavor impacts reactions in quark matter rather differently
 than in a hadron gas. For instance, perturbative QGP interactions at
 relevant temperatures are only mildly affected by the light quark
 mass variations (5-160 MeV).  On the other side, transport properties
 of the heavy $\Omega$ baryons with its mass of 1672 MeV are expected
 to differ completely from the almost massless pions (140 MeV)
 \cite{PRA93}.  Finally, we note that the $\Xi$ slope parameters in Pb
 on Pb collisions are comparable to those in p on W and S on W
 reactions \cite{WA94plb}.  This fact provides strong constraints on
 any reaction dynamics.  While much larger energy densities are surely
 produced in the heavy-ion collisions it does not seem to affect the
 transverse motion of particles like the $\Xi$.  Scenarios in which
 the precursor matter experiences sizable flow before freeze-out of
 the multi-strange hadrons seem therefore to be ruled out by the data
 \cite{footn}.  The conclusion that there can be only a tiny
 collective component in the transverse momentum spectra of the
 multi-strange hadrons is corroborated by an analysis of the chemical
 composition.  The slope parameter $237\pm 24$ MeV of the $\Omega$
 particle is consistent with the ``temperature'' value extracted from
 the particle ratios \cite{bgs}.

\vspace{-1.1cm}
\begin{center}
\begin{figure}[c]
\centerline{\epsfig{figure=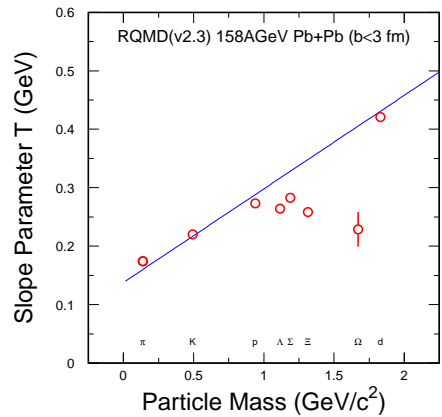,width=7.cm}}
\caption{ The RQMD~2.3 model prediction of the slope parameters. }
\end{figure}
\end{center}
   \vspace{-0.5cm}

 We now turn to the RQMD calculations  in order to go beyond
 qualitative statements. 
The RQMD model \cite{rqmd} provides a 
 microscopic description of
 heavy ion collisions which has been highly
successful in predicting most of the observed features 
over a wide range of conditions. 
 We have generated 1600 events for  Pb(158AGeV) on Pb
 employing  RQMD (version 2.3) and analyzed the final spectra in the  
 same fashion as has been done with the measurements. 
  Before we analyze the
  aspects of the  RQMD evolution  dynamics 
 which are  pertinent for our  discussion 
   we  present the results for the slope parameters
   of the various species  in Fig.~2.  Very 
  good agreement is found  between RQMD predictions and preliminary data. 

\vspace{-.75cm}
\begin{center}
\begin{figure}[c]
\centerline{\epsfig{figure=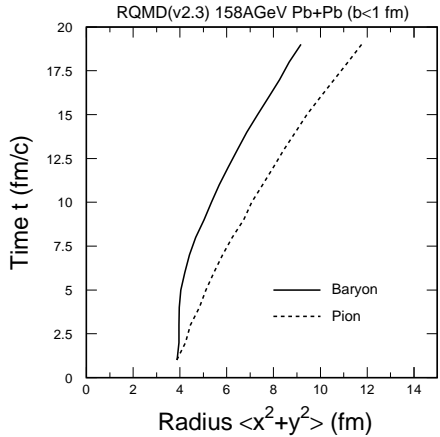,width=7.cm}}
\caption{ Time evolution of the transverse source size for
mid-rapidity baryons and pions. }
\end{figure}
\end{center}
\vspace{-0.5cm}

 The transverse expansion dynamics generated by RQMD may be
 schematically decomposed into two stages.  The pre-equilibrium stage
 is determined by the initial excitation and fragmentation of color
 strings and ropes followed by kinetic equilibration in an ultradense
 hadron gas.  During this stage the effective transverse pressure is
 ultrasoft which is a result of the combined effects from initial
 collision geometry, string dynamics which keeps a memory of the
 original beam axis, finite mean free paths and creation of
 ``resonance matter'' \cite{Sor97}.  According to the RQMD
 calculations this stage lasts for about 5 fm/c in central Pb on Pb
 reactions.  The state when local kinetic equilibrium is finally
 achieved is soon followed by a break-down of equilibrium due to the
 diluteness of the hadron gas and the finite size of the system.  It
 has also been shown that the RQMD evolution of the multi-component
 hadronic fluid is characterized by non-ideal effects, even in the
 dense regime \cite{Sor96}.  The pions accelerate quite easily, their
 motion more or less governed by their own EOS. In contrast, the rarer
 heavy particles cannot keep up with the pion ``fluid'' and are left
 behind.  The developing flow of matter according to the RQMD
 calculations is illustrated by plotting the mean transverse distance
 of hadrons from the center of the collision region (see Fig.~3).  We
 see from Fig.~3 that the baryon matter does not expand at all during
 the first 5 fm/c. Only after the soft stage has elapsed the baryons
 develop collective flow which is reflected in the increase of their
 spatial distribution.  Essentially all of the transverse baryon flow
 is created during the late stage.  Owing to the different transverse
 sizes of the collision region this stage lasts much longer in Pb on
 Pb than in reactions with smaller projectiles and leads to the marked
 difference between momentum spectra for Pb+Pb as compared to p+A and
 S+A reactions.

\vspace{-0.75cm}
\begin{center}
\begin{figure}[c]
\centerline{\epsfig{figure=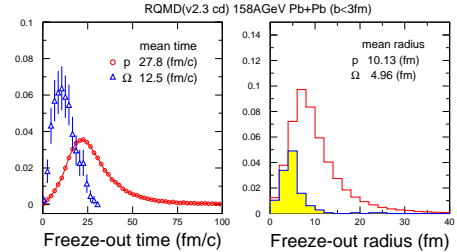,width=7.cm}}
\caption{The  time
and transverse radius distributions of midrapidity  $\Omega$'s and nucleons 
at  freeze-out in central Pb+Pb  collisions at 158AGeV from  RQMD. }   
\end{figure}
\end{center}          
\vspace{-0.75cm}
   
 Why are the multiple strange particles not dragged with the heavy
 particle flow?  Their interactions in the expansion stage are
 dominated by resonance formation.  A good measure of the reaction
 rates are therefore the decay widths of the baryon resonances to
 which the baryons couple. We see from the particle data group tables
 that the decay widths of resonances is a strong function of their
 flavor content \cite{PDG97}. Approximately, the trend is
 0.45:0.62:0.85:1 for $\Omega ^*$:$\Xi ^*$:$Y ^*$:$N^*$.  We expect
 from these numbers that the $\Xi$ collision rates will be suppressed
 by on the order of 30 to 40 percent compared to $\Lambda$'s and
 nucleons.  Furthermore, the $\Omega $ is basically not involved in
 forming these resonances.  A reason is that the $\pi$-$\Omega $
 system does not match any of their flavor quantum
 numbers. Furthermore, $\Omega $+$\eta $ and $\Omega $+$K$ collisions
 are suppressed due to the comparably rarer meson partners.  The
 physics of resonance formation is built into the RQMD approach (for
 details see \cite{rqmd}).  RQMD predicts considerably earlier
 freeze-out of $\Omega $ baryons (and $\Xi$'s) than nucleons (and
 $\Lambda $'s).  The time and spatial distributions of nucleons and
 $\Omega $'s are displayed in Fig.~4.  Most of the $\Omega $'s
 freeze-out between 2 and 8 fm/c while the nucleon freeze-out
 distribution is centered around 20 fm/c.  Note that the freeze-out
 time spectra are characterized by sizable differences between median
 and averages. Very soft collisions between particles of almost equal
 velocities lead to the long tails in the distributions but are of no
 importance for the reaction kinetics.
   
 The corresponding transverse distance distribution which is also
 shown in Fig.~4 reveals that the $\Omega $ source at freeze-out is
 very similar to the initial source.  In contrast, nucleons are
 transported by the collective flow into outer shells.  We can analyze
 the RQMD results concerning the local energy densities at which the
 multi-strange hadrons decouple from the system.  Typically, these
 local energy densities cluster around 1 GeV/fm$^3$. Clearly, this
 sets a lower limit for the densities at which these particles may
 form as quark bound states.
   
 We expect that the RQMD evolution in the late dilute hadron gas stage
 is calculated rather accurately -- under the assumed given initial
 conditions.  For this stage (and only here) our inability to
 calculate nucleus-nucleus reactions based on quantum chromodynamics
 (QCD) can be compensated.  Utilizing the empirical information about
 the interactions between hadrons, kinetic equations can be set up and
 solved like in RQMD and other approaches
 \cite{rqmd},\cite{liko96,cass97,ffm98}.  Theoretical
 justification for semi-classical transport comes essentially from the
 particle-like behavior with the DeBroglie wavelengths being
 typically much smaller than the mean free paths.  Of course, the
 final hadron momentum spectra are a product of the time-integrated
 dynamics starting with the initial interpenetration of the two
 nuclei. The agreement between RQMD and experimental data for the
 final slope parameters implies that the expansion dynamics in the
 first ultradense stage is modeled reasonably well in this approach.
 We thus believe that the ``late'' emergence of the baryon flow holds
 true model-independently.
  
 One might be tempted to look for other explanations of the initial
 ``softness'' than what is provided by the RQMD model. For instance,
 Hung and Shuryak have put forward the idea that the system created in
 Pb+Pb may be close to the so-called softest point of the EOS
 \cite{soft}.  Initial conditions for hydrodynamical calculations may
 be tuned to get good agreement with the non-strange hadron spectra
 measured in Pb on Pb reactions \cite{hu98}.  However, this
 hydrodynamical approach fails to explain the mass dependence of the
 transverse flow, i.e. the data for S projectiles \cite{sh98}. On the
 other side, the RQMD model provides an explanation for the
 systematics from p+A and S+A to Pb+Pb \cite{Sor97}. According to the
 RQMD model the transverse pressure during the early stages is small
 in all these systems, generically a pre-equilibrium signature.

 In summary, we report the results of the analysis of the particle
  transverse momentum distributions from the central 158AGeV Pb+Pb
  collisions. The systematics of the baryon spectra indicates that the
  main component of the baryon flow develops only rather late, after
  most of the multi-strange particles have already frozen out.  We
  infer from the analysis of RQMD results that characteristic energy
  densities at which the multi-strange hadrons freeze out are around 1
  GeV/fm$^3$ for central Pb on Pb collisions.  Presumably, they may be
  formed at even larger densities.


 We are grateful for many enlightening discussions with
 Drs. S. Panitkin, A. Sakaguchi, B. Schlei, E.V. Shuryak, and
 J. Sollfrank.  This research used resources of the National Energy
 Research Scientific Computing Center.  This work has been supported
 by the U.S. Department of Energy under Contract No. DE-AC03-76SF00098
 and W-7405-ENG-36 and National Science Foundation. 

\vspace{-0.25in}


\begin{references}
\bibitem{fn_T} The slope parameter is obtained from fitting the
transverse mass distribution with an exponential function
$exp(-m_T/T)$.
\bibitem{footn} Here as in the following we only refer to the
  transverse directions when we are discussing limits on the
  collective motion.  Undoubtedly, nucleus-nucleus reactions at these
  high energies exhibit a tremendous amount of collective motion in
  the longitudinal direction ``from the beginning'' which is just the
  relic of the initially prepared system.
\end{references}
\end{document}